\documentclass[runningheads]{llncs}
\usepackage[T1]{fontenc}
\usepackage{graphicx}

\usepackage{multirow}
\usepackage{amsmath,bm}
\usepackage{amssymb}
\usepackage{hyperref}
\usepackage{authblk}

\hypersetup{colorlinks}

\begin{document}
\title{Generative Modelling of the Ageing Heart with Cross-Sectional Imaging and Clinical Data}
\titlerunning{Generative Modelling of the Ageing Heart}
%
\author{Mengyun Qiao \inst{1,2} \and Berke Doga Basaran \inst{1,2} \and Huaqi Qiu \inst{1} \and Shuo Wang \inst{7,8} \and Yi Guo \inst{6} \and Yuanyuan Wang \inst{6} \and Paul M. Matthews \inst{3,4} \and Daniel Rueckert \inst{1,5} \and Wenjia Bai \inst{1,2,3}}
\institute{\inst{}Department of Computing, Imperial College London, London, UK
\and \inst{}Data Science Institute, Imperial College London, London, UK
\and \inst{}Department of Brain Sciences, Imperial College London, London, UK \and \inst{}UK Dementia Research Institute, Imperial College London, London, UK
\and \inst{}Klinikum rechts der Isar, Technical University of Munich, Munich, Germany
\and \inst{}Department of Electronic Engineering, Fudan University, Shanghai, China
\and \inst{}Digital Medical Research Center, School of Basic Medical Sciences, Fudan University, Shanghai, China
\and \inst{}Shanghai Key Laboratory of MICCAI, Shanghai, China}

\maketitle              
\begin{abstract}
Cardiovascular disease, the leading cause of death globally, is an age-related disease. Understanding the morphological and functional changes of the heart during ageing is a key scientific question, the answer to which will help us define important risk factors of cardiovascular disease and monitor disease progression. In this work, we propose a novel conditional generative model to describe the changes of 3D anatomy of the heart during ageing. The proposed model is flexible and allows integration of multiple clinical factors (e.g. age, gender) into the generating process. We train the model on a large-scale cross-sectional dataset of cardiac anatomies and evaluate on both cross-sectional and longitudinal datasets. The model demonstrates excellent performance in predicting the longitudinal evolution of the ageing heart and modelling its data distribution. The codes are available at https://github.com/MengyunQ/AgeHeart.

\end{abstract}

\section{Introduction}
Heart ageing is a predominant risk factor of cardiovascular diseases. Understanding how ageing affects the shape and function of the heart is a key scientific question that has received substantial attention \cite{bai2020population,boon2013microrna,eng2016adverse}. Due to the high dimensionality of the 3D cardiac shape data, researchers and clinicians often describe the anatomical shape using global metrics such as the volumes or ejection fraction. However, these metrics cannot reflect detailed information of local shape variations. Describing the high-dimensional spatio-temporal anatomy of the heart and its evolution during ageing is still a challenging problem.

In this work, we propose a novel conditional generative model for the ageing heart, which describes the variations of its 3D cardiac anatomy, as well as its associations with age. The model is trained on a large-scale cross-sectional dataset with both cardiac anatomies and non-imaging clinical information. Once trained, given a cardiac anatomy and a target age, the model can perform counterfactual inference and predict the anatomical appearance of the heart at the target age. By evaluating on both cross-sectional and longitudinal datasets, we demonstrate that the predicted anatomies are highly realistic and consistent with real data distribution. The model has the potential to be applied to downstream tasks for cardiac imaging research, such as for analysing of the ageing impact on the anatomy, synthesising shapes for biomechanical modelling and performing data augmentation.

\subsection{Related Work} 
Numerous efforts have been devoted into conditional generative modelling and synthesis of ageing. In this work, we focus on heart ageing synthesis using conditional generative modelling techniques. Existing literature can be broadly classified into the following two categories:

\subsubsection*{Generative modelling}
The field of generative modelling has made tremendous progress recently, driven by deep learning methods such as variational autoencoders (VAEs) \cite{kingma2013auto,sohn2015learning}, generative adversarial networks (GANs) \cite{mirza2014conditional}, cycle-consistent GAN (CycleGAN) \cite{CycleGAN2017}. Generative models have been widely used in medical imaging. For example, Wang et al. proposed a CycleGAN-based model for cross-domain image generation, which generates pseudo-CT for PET-MR attenuation correction \cite{wang2021dicyc}. Yurt et al. proposed a multi-stream GAN architecture for multi-contrast MRI synthesis \cite{yurt2021mustgan}. Pawlowski et al. formulated a structural causal model with deep learning components for synthesising and counterfactual inference of MNIST and brain MR images \cite{pawlowski2020deep}. 

\subsubsection*{Synthesis of ageing}
Most ageing synthesis works focused on images of human face while some works explored the synthesis of brain MR images. These works investigated different ways of incorporating age information into the generating process. One way is to concatenate age vector with image feature vector to learn a joint distribution of age and image appearance in face ageing \cite{antipov2017face,zhang2017age} or brain ageing \cite{xia2019consistent}. Another way is to use a pre-trained age regression network, which provides guidance in age-related latent code generation \cite{alaluf2021only,wang2018face}. Some works introduced an age estimation loss accounting for age distribution \cite{huang2020pfa,or2020lifespan}. In \cite{makhmudkhujaev2021re}, high-order interactions between the given identity and target age were explored to learn personalized age features. Although these methods are not designed for cardiac imaging, they provide valuable insights for modelling the ageing heart.

\subsection{Contributions}
The contributions of this work are three-fold: 1) We investigate the challenging problem of heart ageing synthesis, where both the structural variation and functional variation (anatomies in different time frames) need to be modelled. To this end, we develop a novel model which consists of an anatomy encoder and a condition mapping network that disentangles age and spatial-temporal shape information in the generating process; 2) We utilise multi-modal information including both imaging data and non-imaging clinical data so that the generative model can account for the impact of multiple clinical factors on the ageing heart; 3) We train the generative model using a large-scale cross-sectional datasets and demonstrate its performance quantitatively on both cross-sectional and longitudinal datasets. To the best of our knowledge, this is the first work to investigate generative modelling for ageing heart synthesis.

\section{Methods}
\subsection{Problem Formulation}
\begin{figure}
\centering
\includegraphics[width=0.95\textwidth]{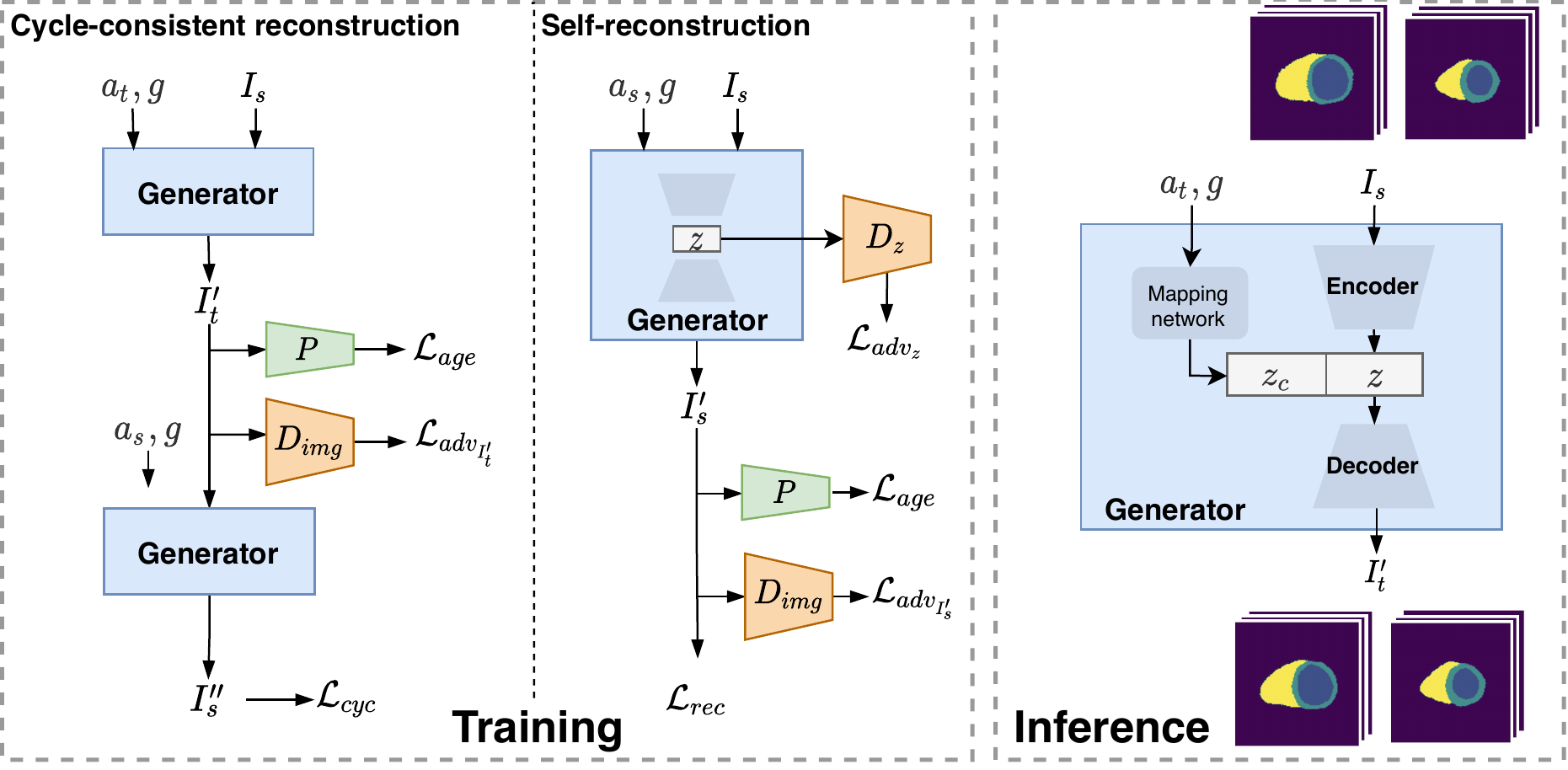}
\caption{The proposed generative model for the ageing heart. \textbf{Training:} The training scheme includes two parts: cycle-consistent reconstruction, which takes source image $I_s$ at source age $a_s$ as input, synthesise $I'_t$ at target age $a_t$ and then back to $I''_s$ at age $a_s$; self-reconstruction, which takes source image $I_s$ as input and reconstruct $I'_s$ at the same age. \textbf{Inference:} The Generator takes input image $I_s$ at source age $a_s$, together with target age $a_t$ and gender $g$, and generates the image $I_t$ at target age $a_t$. Please refer to the text for detail.} \label{figx_network}
\end{figure}

Fig.~\ref{figx_network} illustrates the proposed generative model. At the inference stage (Fig.~\ref{figx_network} right), given a source cardiac anatomy image $I_s$ with its clinical information (source age: $a_s$, gender: $g$) and the target age $a_t$, the network synthesises the cardiac anatomy $I'_t = G(I_s,a_t,g)$ conditioned on the target age $a_t$. Thus, the distribution of the synthetic anatomy approximates the distribution of the real data at the target age while maintaining subject-specific structure in the ageing process. Our model utilises a heart anatomy generator $G$, which consists of an encoder $E$, an anatomy decoder $D$ and a condition mapping network $M$. During the training stage, the generator $G$ learns the evolution of the anatomy $I_s$ from the source age $a_s$ to the anatomy $I_t$ at target age $a_t$ and vice versa in a cyclic manner. 

\subsection{Conditional Generative Modelling}
\subsubsection{Clinical condition incorporation}
We incorporate two major clinical conditions, age and gender, into the generative model. The age space $A$ is represented using a $m \times 1$ categorical vector, where $m$ denotes the number of age groups. For age group $i$, an age vector $a_i\in A$ is generated:
\begin{equation}\label{noise}
a_i=\mathrm{A_i} + \varepsilon, \varepsilon \sim \mathcal {N}(0,\ {\sigma}^2)
\end{equation}
where $\mathrm{A_i}$ is a one-hot encoded vector that contains one at the $i$th-element and zeros elsewhere and $\varepsilon$ is random noise sampled from a prior distribution. The age vector $a_i$ is concatenated with a one-hot gender vector $g$ to form the clinical condition $c$.

\paragraph{Condition mapping network}
Inspired by \cite{karras2020training}, we construct a condition mapping network $M(a,g)$ using a multi-layer perception (MLP). It embeds the input clinical condition $c$ including age $a$ and gender $g$ to latent vector $z_c$ in the conditional latent space. This latent representation integrates different clinical factors and enables exploration across the condition space.

\paragraph{Age predictor}
We construct an age predictor $P$ to help the generative model focus on the age information in learning. $P$ takes a cardiac anatomy as input and predicts its age. It is applied only at the training stage for both cycle-consistent reconstruction and self-reconstruction as shown in Fig.~\ref{figx_network}. We impose a distance loss between the predicted age $a'_t=P(I'_t)$ and target age $a_t$, as well as between $a'_s = P(I'_s)$ and $a_s$ to guide the age prediction. The age predictor is implemented as a six-layer 3D convolutional network, followed by a fully connected layer to produce an age vector.

\subsubsection{Anatomy generator}
The input to the generator $G$ includes the source anatomy $I_s$ and the condition code $z_c$ generated from the condition mapping network $M(a,g)$. The generator follows an encoder-decoder structure, where the encoder $E$ maps the input $I_s$ into a subject-specific latent code $z$, the decoder concatenates $z$ with the clinical condition code $z_c$ and generates an output anatomy $I'_t$. The generation is described by,
\begin{equation}\label{eq-gen}
I'_t = G(I_s,a_s,g) = D(E(I_s), M(a_s,g))
\end{equation}

We assume that the encoder $E(I_s)$ preserves the high-level subject-specific feature of the input anatomy $I_s$ and the decoder $D$ utilises this information as well as the clinical information to generate the anatomy $I'_t$. Adopting a cyclic design \cite{CycleGAN2017}, we also generate an image $I''_s=G(I'_t,a_s,g)$ that maps the generated the image $I'_t$ back to the source age $a_s$. This cycle-consistent generation is only applied in training process.

\subsubsection{Discriminators}
Two discriminator networks are imposed on the latent code $z$ and the generated anatomy. $D_z$ is designed to discriminate the latent code $z=E(I)$ by training $z$ to be uniformly distributed. Simultaneously, $E$ will be trained to compute $z$ to fool $D_z$. Such an adversarial process forces the distribution of $z$ to gradually approach the prior, which is the uniform distribution. Another discriminator $D_{img}$ forces the generator to generate realistic cardiac anatomies.

\subsection{Training Scheme}
An overview of the training scheme is shown in Fig.~\ref{figx_network}. To generate realistic anatomies while modelling smooth continuous ageing, we use a multi-task loss function which combines cyclic reconstruction losses $L_{rec}$ and $L_{cyc}$, adversarial losses $L_{adv_{I'_t}},L_{adv_{I'_s}}$ for the anatomy discriminator $D_{img}$, an adversarial loss $L_{adv_z}$ for the latent code discriminator $D_z$ and an age loss $L_{age}$ for the age predictor $P$. For heart ageing synthesis, the source anatomy $I_s$ and the generated anatomy $I'_s = G(I_s, a_s, g)$ at the same age $a_s$ are expected to be similar. A self-reconstruction loss between source image $I_s$ and reconstruction image $I'_s$ is applied to learn the identity generation. In addition, we employ the cycle consistency loss \cite{CycleGAN2017} between $I_s$ and $I''_s = G(I'_t, a_s, g)$ for a consistent reconstruction from age $a_s$ to age $a_t$ and back to age $a_s$. L1 loss is used for reconstruction:
\begin{equation}
\centering
\mathcal{L}_{rec}(G)=\left \| I_s-I'_s \right \|_1,\mathcal{L}_{cyc}(G)=\left \| I_s-I''_s \right  \|_1
\label{eq:loss_recon}
\end{equation}

The generated images $I'_t$, $I'_s$ are enforced to the target age space by minimizing the distance between the age predictor outputs $P(I'_t)$, $P(I'_s)$ and the age vectors $a_t$, $a_s$. A cross-entropy (CE) age loss is defined as,
\begin{equation}
\centering
\mathcal{L}_{age}(G)=\left \| a_t-P(I'_t) \right  \|_{CE(a_t,P(I'_t))}+\left \| a_s-P(I'_s) \right  \|_{CE}
\label{eq:loss_age}
\end{equation}

An adversarial loss $L_{adv_z}$ is used to impose an uniform distribution on the latent code $z=E(I_s)$:
\begin{equation}
\mathcal{L}_{adv_z}(E,D)= \mathbb{E}_{z^*}\left[ \log D_z(z^*)  \right] + \mathbb{E}_{I_s}\left[ \log (1-D_z(E(I_s)))  \right]
\end{equation}
where $z^*$ denote random samples from a uniformed prior distribution.

In addition, two adversarial losses conditioned on the source and target ages of the real and synthetic anatomies are introduced, respectively:
\begin{equation}\label{advz}
\mathcal{L}_{adv_{I'_t}}(G,D)= \mathbb{E}_{I_s,a_s}\left[ \log D_{img}(I_
s,a_s)  \right] + \mathbb{E}_{I'_t,a_t}\left[ \log (1-D_{img}(I'_t))  \right]
\end{equation}
\begin{equation}\label{advimg}
\mathcal{L}_{adv_{I'_s}}(G,D)= \mathbb{E}_{I_s,a_s}\left[ \log D_{img}(I_s,a_s)  \right] + \mathbb{E}_{I_s,a_s}\left[ \log (1-D_{img}(I'_s))  \right]
\end{equation}
The adversarial losses presented in Eq. \ref{advz} and Eq. \ref{advimg}, minimizing the distance between the input and output images, forces the output anatomies to be close to the real ones.

Overall, the optimisation is formulated as an adversarial training process,
\begin{equation}\label{optimisation}
\begin{split}
\underset{G,E}{\min}\ \underset{D_z,D_{img}}{\max}\ \lambda _0 \mathcal{L}_{rec}(G)+\lambda _1 \mathcal{L}_{cyc}(G)+\lambda _2 \mathcal{L}_{age}(G)+ \\
  \lambda _3 \mathcal{L}_{adv_z}(E,D)+\lambda _4 \mathcal{L}_{adv_{I'_t}}(G,D)+\lambda _5 \mathcal{L}_{adv_{I'_s}}(G,D)
\end{split}
\end{equation}
where the $\lambda$'s are tunable hyperparameters weighting the loss terms.

\section{Experiments}
\subsection{Datasets} \label{data}
\noindent\textbf{Cross-sectional dataset} Short-axis cardiac images at the end-diastolic (ED) and end-systolic (ES) frames of 12,600 subjects from 44.6 to 82.3 years old, were obtained from the UK Biobank and split into training ($n = 11,340$) and test ($n = 1,260$) sets. The age is represented as seven categories ($m=7$) with interval of five years: 44–50, 50-55, 55-60, 60-65, 65-70, 70-75 and 75-83. Most of datasets are from healthy volunteers and about 5$\%$-6$\%$ have the cardiovascular diseases (CVD), which we would take into consideration in the future.

\noindent\textbf{Longitudinal dataset}
A longitudinal dataset of 639 subjects from the UK Biobank is used, in which each subject undergoes imaging at two time points. The age ranges from 46.6 to 79.8 years old at the first imaging and 51.3 to 81.9 years at the re-imaging, with a median time gap of 3.2 years. The image resolution and size are the same as the cross-sectional dataset. All evaluations are performed on ED and ES frames of cardiac sequences.

\noindent\textbf{Preprocessing}
For both datasets, 3D cardiac anatomies at ED and ES frames are extracted from cardiac MR images using a publicly available segmentation network \cite{bai2018automated}, then upsampled using a publicly available super-resolution model \cite{wang2021joint} followed by manual quality control. Subsequently, affine registration is performed to align all cardiac anatomies to the same orientation. The 3D cardiac anatomies are of an isotropic resolution of $1.8\times 1.8\times 2$ mm$^3$ and of size $128\times128\times64$ voxels.

\subsection{Experimental Setup}
\subsubsection*{Implementation details}
The encoder $E$ consisted of five 3D convolutional layers and one flatten layer, outputting the latent code $z$. The decoder $D$ consisted of one flatten layer and five 3D transposed convolution layer. The transposed convolution in encoder and decoder used a kernel size of $4\times4\times4$. All intermediate layers of each block use the ReLU activation function. The dimension of the latent variables $z$ and $z_c$ are both 32. The anatomy segmentations are transferred into one-hot map, and the output of $E$($I$) is restricted to $[-1,1]$ using the hyperbolic tangent activation function. For optimisation, the Adam optimizer \cite{kingma2014adam} is used with learning rate of $2\cdot 10^{-4}$ and weigh decay of $1\cdot 10^{-5}$. We set $\sigma = 0.02$ in Eq. \ref{noise}, $\lambda _0=1$, $\lambda _1=0.1$, $\lambda _2=0.01$, $\lambda _3=0.1$ and $\lambda _4=1$ in Eq.~\ref{optimisation}. The model was implemented using PyTorch \cite{paszke2019pytorch}. At the inference stage, only the generator $G$ is active, containing $E$, $D$ and $M$ as described in Eq.~\ref{eq-gen}, while the other parts are not used. Our code will be made publicly available.

\subsubsection*{Baseline methods}
Two ageing synthesis methods, CAAE \cite{zhang2017age} and a modified version of Lifespan \cite{or2020lifespan}, are used as baselines. Since the original codes were developed for 2D face image synthesis, we re-implemented all the codes for 3D cardiac data synthesis. For Lifespan, we replaced the modulated convolution layers with 3D convolution layers to save GPU memory for 3D data.

\subsection{Experiments and Results}
\subsubsection{Heart ageing synthesis}
\begin{figure}[ht]
\centering
\includegraphics[width=0.85\textwidth]{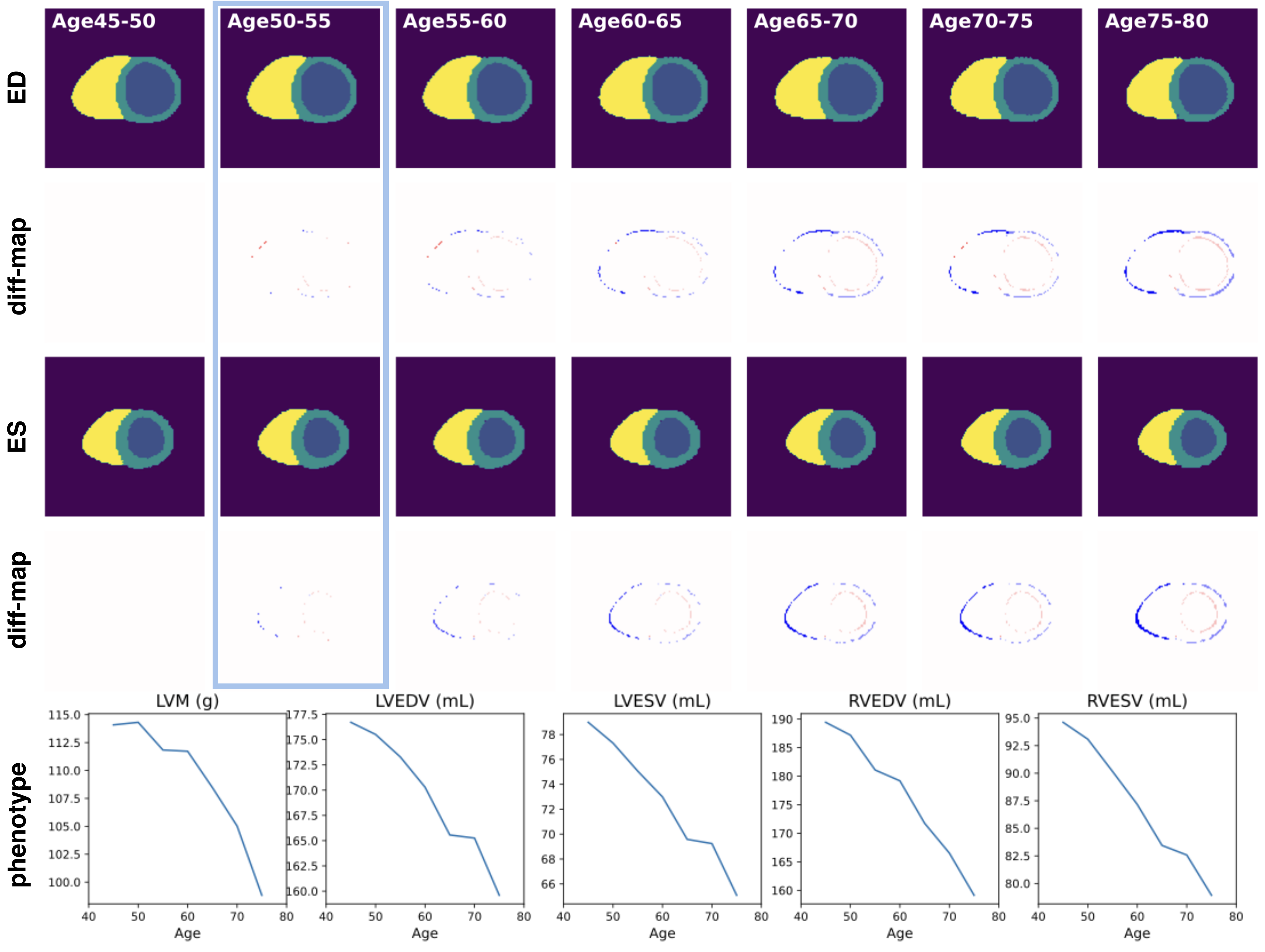}
\caption{An example of the synthetic ageing heart. The first and third rows show the cardiac anatomies at ED and ES frames, in which the blue rectangle denotes the original input anatomy of a 50-55 year old female and the other columns denote the synthetic anatomies at different ages. The second and fourth rows show the difference maps between an aged heart and the baseline anatomy at 45-50 year old. The fifth shows the predicted evolution of clinical measures including LVM, LVEDV, LVESV, RVEDV and RVESV during ageing.}
\label{figx-age}
\end{figure}

For each subject in the UK Biobank test set, we synthesise a series of anatomies for the same heart at age groups from 40 to 80 with interval of 5 years old. For example, in Fig.~\ref{figx-age}, the cardiac anatomy of a 50-55 years old female is taken as input and the anatomies of this heart at other ages are predicted using the proposed generative model. From the generated anatomies, we also derive clinical measures, including the left ventricular myocardial mass (LVM), LV end-diastolic volume (LVEDV), LV end-systolic volume (LVESV), RV end-diastolic volume (RVEDV) and RV end-systolic volume (RVESV). The bottom of Fig.~\ref{figx-age} illustrates the trends of these clinical measures during heart ageing synthesis. Consistent with the literature \cite{eng2016adverse}, we observe a decreasing trend for LV or RV volumes. It demonstrates our model captures the relation between cardiac anatomical structure and age.
\subsubsection{Distribution similarity}
Based the synthetic anatomies, we calculate the probability distribution $P_c(a)$ of each clinical measure $c$ against age $a$ and compare it to the probability distribution of the real data $Q_c(a)$. Here, $c$ denotes one of the five clinical measures, e.g. LVM. The distance between $P_c(a)$ and $Q_c(a)$ is evaluated in terms of the Kullback–Leibler divergence (KL) and Wasserstein distance (WD). Table~\ref{tab1} compares the distribution similarities using different generative models. It shows the proposed method achieves a higher distribution similarity, compared to state-of-the-art ageing synthesis models.

\begin{table}[!h]
\caption{The distribution similarity between synthetic and real data. The smaller the KL or WD distance, the higher the similarity.}\label{tab1}
\centering
\resizebox{\textwidth}{!}{
\begin{tabular}{ccccccccccc}
\hline
&
  \multicolumn{2}{c}{LVM} &
  \multicolumn{2}{c}{LVEDV} &
  \multicolumn{2}{c}{LVESV} &
  \multicolumn{2}{c}{RVEDV} &
  \multicolumn{2}{c}{RVESV} \\ \cline{2-11}
                               & KL     & WD      & KL    & WD       & KL     & WD     & KL     & WD      & KL     & WD  \\ \hline
CAAE \cite{zhang2017age}       & 0.0266 & 19.8875 &0.0355 & \textbf{15.6661}  & 0.0737 & 9.8066 & 0.0343 & \textbf{18.7905} & 0.0467 & 11.9521 \\
Lifespan \cite{or2020lifespan} & 0.0253 & 18.4733 &0.0349 & 15.8506  & 0.0703 & 9.8091 & 0.0322 & 20.0034 & 0.0442 & 12.1956 \\
Proposed &
  \textbf{0.0248} &
  \textbf{15.2829} &
  \textbf{0.0334} &
  15.7215 &
  \textbf{0.0675} &
  \textbf{9.5658} &
  \textbf{0.0318} &
  19.1561 &
  \textbf{0.0428} &
  \textbf{10.8607} \\ \hline
\end{tabular}
}
\end{table}

\subsubsection{Longitudinal prediction}
Using the repeated imaging scans from UK Biobank longitudinal dataset, we evaluate the predictive performance of the model. Given the anatomy at the first time point, the anatomy at the second time point is predicted and compared to the ground truth in terms of Dice metric, Hausdorff distance (HD) and average symmetric surface distance (ASSD), reported in Table~\ref{table-longitudinal}. It shows that the proposed method achieves a good performance in prediction comparable to or better than other competing methods.
\begin{table}[!h]
\caption{The prediction performance on the UK Biobank longitudinal dataset. The higher Dice or lower HD and ASSD, the better the prediction.}
\label{table-longitudinal}
\centering
\resizebox{\textwidth}{!}{
\begin{tabular}{ccccccc}
\hline
 & \multicolumn{3}{c}{End-diastolic anatomy}     & \multicolumn{3}{c}{End-systolic anatomy}               \\ \cline{2-7} 
                        & Dice         & HD            & ASSD         & Dice         & HD            & ASSD                  \\ \hline
CAAE \cite{zhang2017age}                   & 0.727(0.057) & 30.431(6.058) & 2.777(0.788) & 0.769(0.070) & 15.904(6.766) & 2.528(1.054)          \\
Lifespan \cite{or2020lifespan}         & 0.757(0.064) & 29.935(5.988) & 2.784(0.806) & 0.774(0.072) & 16.023(6.527) & \textbf{2.490(1.041)} \\
Proposed & \textbf{0.761(0.066)} & \textbf{27.281(7.436)} & \textbf{2.695(0.835)} & \textbf{0.775(0.073)} & \textbf{14.789(7.224)} & 2.524(1.073) \\ \hline
\end{tabular}
}
\end{table}
\section{Conclusion}
To conclude, we propose a novel generative model for the ageing heart anatomy that allows preserving the identity of the heart while changing its characteristics across different age groups. The quantitative results on both cross-sectional and longitudinal datasets demonstrate the method achieves highly realistic synthesis and longitudinal prediction of cardiac anatomies, which are consistent with real data distributions. 


\subsubsection{Acknowledgements}
This work was supported by EPSRC SmartHeart Grant (EP/P001009/1) and DeepGeM Grant (EP/W01842X/1). The research was conducted using the UK Biobank Resource under Application Number 18545. We wish to thank all UK Biobank participants and staff.

%
%
\bibliographystyle{splncs04}
\bibliography{reference.bib}

\newpage
\pagestyle{empty}
\setcounter{secnumdepth}{0}

\setcounter{figure}{0}
\setcounter{table}{0}
\renewcommand{\thefigure}{A\arabic{figure}}
\renewcommand{\thetable}{A\arabic{table}}

\end{document}